\documentclass[12pt,a4paper]{article}
\usepackage{thophys}
\usepackage{graphics}
\usepackage{amsmath,amssymb}
\DeclareMathOperator{\tr}{tr}
\DeclareMathOperator{\astcomma}{\stackrel{\ast}{,}}
\newcommand{\ii}{\mathrm{i}}
\newcommand{\ee}{\mathrm{e}}

\allowdisplaybreaks
\makeindex
\begin{document}
\title{%
  \mbox{The Noncommutative Standard Model at~$\mathcal{O}(\theta^2)$}}
\author{%
  Ana Alboteanu%
    \thanks{e-mail: \texttt{aamaria@physik.uni-wuerzburg.de}}\\
  Thorsten Ohl%
    \thanks{e-mail: \texttt{ohl@physik.uni-wuerzburg.de}}\\
  Reinhold R\"uckl%
    \thanks{e-mail: \texttt{rueckl@physik.uni-wuerzburg.de}}\\
  \hfil\\
    Institut f\"ur Theoretische Physik und Astrophysik\\
    Universit\"at W\"urzburg\\
    Am Hubland, 97074 W\"urzburg, Germany}
\date{July 2007}
\maketitle
\begin{abstract}
  We derive the most general Seiberg-Witten maps for noncommutative
  gauge theories in second order of the noncommutative parameter~$\theta$.  
  Our results reveal the existence of 
  more ambiguities than previously known.  In particular, we
  demonstrate that some of these ambiguities enter observables like
  scattering cross sections and enlarge the parameter space of the
  noncommutative standard model beyond~$\mathcal{O}(\theta)$.
\end{abstract}
\newpage
\section{Introduction}

With the start of data taking at the Large Hadron Collider~(LHC),
particle physics will, for the first time, directly probe the Tera Scale,
i.\,e.~the scale of electroweak symmetry breaking according to the
Standard Model~(SM), around~$1\,\mathrm{TeV}$.  While the~SM has been
confirmed experimentally to be a very precise effective description of
the physics below the Tera Scale, there are many serious contenders for
the more fundamental theory beneath the~SM.

For quite some time, superstring theory has been a leading
candidate for the fundamental theory unifying all known interactions.
There are certain solutions of superstring theory with additional spatial dimensions, 
where the characteristic string scales are low enough to allow
experimental tests at the LHC and the planned International $e^+e^-$
Linear Collider~(ILC). One spectacular prediction~\cite{Seiberg:1999vs} 
of superstring theory is the emergence of a noncommutative~(NC) 
structure of spacetime 
at a scale~$\Lambda_{\text{NC}}$ associated with nonvanishing 
commutators
\begin{equation}
\label{eq:NC-geometry}
  \left[ x_\mu , x_\nu \right] = \ii \theta_{\mu\nu}
     = \ii \frac{1}{\Lambda_{\text{NC}}^2}C_{\mu\nu}
\end{equation}
of spacetime coordinates that 
correspond to oriented minimal resolvable areas of
size~$\mathcal{O}(\Lambda_{\text{NC}}^{-2})$.  While a
nonvanishing commutator like~(\ref{eq:NC-geometry}) had been proposed much
earlier~\cite{Snyder} as a regulator of divergencies in Quantum Field
Theory~(QFT) and Quantum Gravity, the observation
of~\cite{Seiberg:1999vs} caused intense renewed interest in QFT on NC
spacetimes (NCQFT).

The commutator~(\ref{eq:NC-geometry}) can be conveniently realized on
a \emph{commuting} spacetime by replacing all products of functions by
Moyal-Weyl $\ast$-products
\begin{equation}
  (f\ast g)(x) =
    f(x) e^{ \frac{\ii}{2}\overleftarrow{\partial^\mu}
               \theta_{\mu\nu}
               \overrightarrow{\partial^\nu}} g(x)\,.
\end{equation}
A prescription for constructing arbitrary gauge theories on a NC
spacetime was presented in~\cite{Seiberg:1999vs}.  These so-called
Seiberg-Witten Maps~(SWM) realize NC gauge transformations in the NC
theory as ordinary commutative gauge transformations on an effective 
commutative gauge theory.  
By going to the enveloping algebra of the Lie algebra of a given 
gauge group, this approach~\cite{Jurco:2001rq} circumvents obstructions
like charge quantization in~$U(1)$ gauge theories and the
prohibition of~$SU(N)$ gauge groups in the earlier attempts.

In particular, this prescription allowed the construction of
the so-called 
Noncommutative Standard Model~(NCSM)~\cite{NCSM} as an
anomaly-free~\cite{Anomaly-Freedom} canonical
NC extension of the SM without having to introduce additional 
particles\footnote{Other
  constructions of NC extensions of the SM start from
  a~$U(3) \otimes U(2) \otimes U(1)$ gauge theory
  and subsequently break the extraneous symmetries, introducing
  additional particles that must be removed from the observable
  spectrum~\cite{Chaichian:2001py}.}.
In the first order of an expansion in~$\theta$, one has 
only three new bounded parameters that depend on the choice of the
representation of the enveloping algebra of the SM Lie
algebra and describe new couplings among gauge bosons~\cite{NCSM}.
These couplings vanish in the minimal NCSM, where the enveloping
algebra is realized by matrices acting in the vector space of the adjoint
representation.
Furthermore, the bosonic sector of the minimal NCSM was shown to be
renormalizable at one loop~\cite{NCSM-renormalization:bosons}, where
all counter terms can be expressed through the usual field strength and
coupling constant renormalizations.  Also the nonminimal NCSM is 
renormalizable at one-loop, if a \emph{finite} gauge
invariant $\mathcal{O}(\theta)$-term is added to the
action~\cite{NCSM-renormalization:bosons}.  Finally, the fermionic
sector can be shown to require a finite number of gauge invariant
four-fermion operators as additional counter
terms in one loop order~\cite{NCSM-renormalization:fermions}.
In euclidean NC space, the renormalizability of scalar and gauge
models has been shown to all orders in~$\theta$~\cite{Renormalizability}.

Using the effective theory in the first order of the
$\theta$-expansion, several phenomenological studies were performed
for past, present, and future
colliders~\cite{NCSM-Pheno,Ohl/Reuter:2004:NCPC,Alboteanu/Ohl/Rueckl:2006}.
In a preceding paper~\cite{Alboteanu/Ohl/Rueckl:2006}, we have studied
the associated production of photons and $Z$-bosons at hadron
colliders (Tevatron and LHC) showing that at the LHC one can reach a
noncommutativity scale~$\Lambda_{\text{NC}}$ slightly
above~$1\,\mathrm{TeV}$~\cite{Alboteanu/Ohl/Rueckl:2006}.
Moreover, we have found that it is necessary to go beyond the
first order in~$\theta$, because of significant contributions from
partonic center of mass energies exceeding the noncommutativity scale 
that can actually be probed.

While it is possible in simple cases to derive expressions for
families of SWM to all orders
in~$\theta$~\cite{Barnich:2003wq,Zeiner/Rauh:Theses,Ohl/Rauh/Rueckl/Zeiner:2007:AOSWM},
an explicit parameterization of the most general solution has not been
given.  Therefore, we start in this
paper by constructing the most general SWM for the NCSM in the second
order of the $\theta$-expansion.  The importance of this systematic
approach is stressed \textit{a posteriori} by discovering
ambiguities in the SWM that have been missed in
earlier~$\mathcal{O}(\theta^2)$ constructions of
SWM~\cite{Jurco:2001rq,Moller:2004qq}. While these authors expected
all ambiguities to cancel in observable quantities,
we find that they do \emph{not}.  
In fact, using $e^+e^-\to\gamma\gamma$ as an example, we will calculate 
the ambiguity in the corresponding scattering amplitude explicitly.

The outline of this paper is as follows. In section~\ref{sec:SWM} we
derive the general SWM up to second order in the
noncommutativity $\theta$.  Particular emphasis will be given to the
ambiguities resulting from the homogeneous solutions of the gauge
equivalence equations.  Furthermore, the Lagrangian and the Feynman
rules of the neutral current sector of the NCSM are constructed in 
section~\ref{sec:Feynman}. Section~\ref{sec:ambiguities} presents our
analysis of the impact of the SWM ambiguities on physical observables.  
We will demonstrate by an explicit calculation
of~$e^+e^- \to \gamma \gamma$ that not all ambiguities cancel in
the~$\mathcal{O}(\theta^2)$ contribution to the cross section. 
In section~\ref{sec:summary} we conclude with a brief summary.
All expressions that are needed for the main results of this 
paper will be given in full, either in the main text or in the appendices.  
Complete expressions in second order in $\theta$ that are too lengthy 
to be included in this paper can be found in the appendix 
of~\cite{Alboteanu:2007}.

\section{Seiberg-Witten Maps}
\label{sec:SWM}

The purpose of the SWM is to realize noncommutative gauge
transformations by representations of the enveloping algebra using
nonlinear functions of ordinary, commutative fields that reside 
in representations of the given Lie algebra.  
This requirement is expressed by a set of 
so-called gauge equivalence equations for noncommutative 
gauge fields~$\hat A_\mu(A, \theta)$, 
gauge parameters\footnote{The gauge
  parameters~$\hat\lambda(\alpha,A,\theta)$ appear in the Lagrangian
  in the guise of Faddeev-Popov ghosts, if the gauge fixing is
  performed before application of the SWM.}~$\hat\lambda(\alpha,A,\theta)$ 
and matter fields~$\hat \psi (\psi, A, \theta)$ as functions of
the commutative gauge fields~$A_\mu$,
gauge parameters~$\alpha$ and 
matter fields~$\psi$.
General SWM are defined as solutions
of the gauge equivalence equations\footnote{The gauge equivalence
equations~(\ref{eq:SWM-condition}) could be relaxed by demanding that
the two sides of equation~(\ref{eq:SWM-condition}) lie in the same
gauge orbit, but are not identical~\cite{Asakawa:1999cu}.  Since by
construction the corresponding ambiguities must cancel in the gauge
invariant Yang-Mills action to all orders in $\theta$, we can
ignore them in the rest of this paper.}:
\begin{subequations}
\label{eq:SWM-condition}
\begin{align}
\label{eq:SWM-condition-A}
  \hat A_\mu (A, \theta)
     &\to \ee^{\ii\hat\lambda(\alpha,A,\theta)\ast}
            \left(\hat A_\mu (A, \theta)+\ii\partial_\mu\right)
          \ee^{-\ii\hat\lambda(\alpha,A,\theta)\ast}
      \stackrel{!}{=} \hat A_\mu (A', \theta) \\
\label{eq:SWM-condition-psi}
  \hat \psi (\psi, A, \theta)
     &\to \ee^{\ii\hat\lambda(\alpha,A,\theta)\ast} 
        \hat \psi (\psi, A, \theta)
           \stackrel{!}{=} \hat \psi (\psi', A', \theta)\,,
\end{align}
\end{subequations}
where $A_\mu$ and $\psi$ transform as usual:
\begin{subequations}
\label{eq:gauge-transformations}
\begin{align}
  A_\mu &\to A_\mu' = \ee^{\ii\alpha} 
     \left( A_\mu+\ii\partial_\mu\right) \ee^{-\ii\alpha}\\
  \psi &\to \psi'   = \ee^{\ii\alpha} \psi\,.
\end{align}
\end{subequations}
Here, we have used the notation $A_\mu= A^a_\mu T^a$ and
$\alpha=\alpha^a T^a$, $T^a$ being the generators of the gauge group. 
In practice, the gauge equivalence equations~(\ref{eq:SWM-condition}) 
can be solved order by order in an expansion in~$\theta$.

\subsection{Field Redefinitions vs.~SWM Ambiguities}
\label{sec:field-redefinitions}

The physical predictions of QFT, in particular
the on-shell $S$-matrix elements and scattering cross sections, do not
depend on the choice of interpolating
fields~\cite{Haag/Ruelle,CCWZ,Weinberg:1978kz}.  In fact, any two
theories which are related by non-singular local field redefinitions
\begin{subequations}
\label{eq:reparametrization}
\begin{equation}
  \Phi \leftrightarrow \Phi'(\Phi)
     \;\text{with}\;
  \frac{\partial\Phi'}{\partial\Phi}\biggr|_{\Phi=0}=\mathbf{1}
\end{equation}
and the corresponding change in the Lagrangian
\begin{equation}
  \mathcal{L}(\Phi) \leftrightarrow \mathcal{L}'(\Phi') 
    = \mathcal{L}(\Phi(\Phi'))
\end{equation}
\end{subequations}
will predict identical scattering cross sections.  This
reparametrization invariance can be proven both in
axiomatic QFT~\cite{Haag/Ruelle} and in perturbation
theory for effective QFT~\cite{Weinberg:1978kz}.  It provides
the basis for the application of the powerful and now ubiquitous
methods of effective QFT to elementary particle
physics phenomenology~\cite{CCWZ}.   The fact that the
reparametrization~(\ref{eq:reparametrization}) corresponds to a change
of integration variables in the path integral, provides a particularly
intuitive proof of the invariance.

Since the SWM
\begin{equation}
\label{eq:SWM}
  \begin{pmatrix}
    \alpha \\
    A_\mu \\
    \psi
  \end{pmatrix} \to
  \begin{pmatrix}
    \hat\lambda(\alpha,A,\theta) \\
    \hat A_\mu(A,\theta) \\
    \hat \psi(\psi,A,\theta)
  \end{pmatrix}
\end{equation}
appear to be just a reparametrization 
like~(\ref{eq:reparametrization}), one might expect that its
application has no effect on the calculation of observables. 
For a noncommutative $U(1)$ gauge theory with unit
charge~\cite{Wulkenhaar:SWM-theta-expansion} it can be shown 
that this is indeed the case.
However, in theories with
multiple $U(1)$ charges and $SU(N)$ gauge groups,
the enveloping algebra is strictly larger than the Lie algebra and the
SWM~(\ref{eq:SWM}) must therefore be singular.  The NCSM is a prominent
representative of such theories, whence we must expect nontrivial effects
of the SWM on observables, as can readily be seen 
from~\cite{NCSM-Pheno,Ohl/Reuter:2004:NCPC,Alboteanu/Ohl/Rueckl:2006}.
It was noted from early on, that the solutions
of~(\ref{eq:SWM-condition}) are not
unique~\cite{Jurco:2001rq,Moller:2004qq}.  Consequently, the
construction of NC extensions of the SM via SWM is a priori not
unique as well. Only those ambiguities that correspond to non-singular
reparametrization like~(\ref{eq:reparametrization}) are guaranteed to
cancel in observables.  This leaves us with the crucial problem
of genuine ambiguities in physical quantities.

To first order in~$\theta$ it has been shown by
explicit calculations that all ambiguities cancel in on-shell 
scattering amplitudes.  In turn, one
can use these cancellations as a powerful consistency check for
numerical calculations of cross
sections~\cite{Alboteanu/Ohl/Rueckl:2006}.  However, there is no 
general theorem or physical argument that implies the cancellation 
of all ambiguities in the SWM to all orders in the $\theta$-expansion.
Below we will identify ambiguities in the second order SWM that
affect cross sections.  In an
examination of the effective action of
$\mathcal{O}(\theta)$-NCQED~\cite{Wulkenhaar:SWM-theta-expansion} 
showing that the SWM in first order corresponds to a field redefinition, 
it was already conjectured that this is not the case in second order.  
Our results provide a proof for this conjecture from a completely 
different angle.  Furthermore, some of the
ambiguities in second order are discussed in~\cite{Moller:2004qq}, 
but no claim is made of completeness.  
In fact, we have found additional ambiguities.

A recursive solution of the gauge equivalence
equations~(\ref{eq:SWM-condition}) to all orders in~$\theta$,
including ambiguities, has been given in~\cite{Barnich:2003wq}.
Unfortunately, the ambiguities are not given as an explicit function
of an independent set of parameters and it is not straightforward to
search for observable effects in the NCSM in this form.  Thus we do
not rely on~\cite{Barnich:2003wq} for a discussion of the general
solution of~(\ref{eq:SWM-condition}) and derive the higher order terms
anew.

Still, the existence of ambiguities with observable effects does
not render the NCSM proposed in~\cite{NCSM} unphysical and 
thus rather useless.
It merely adds more free
parameters, not unlike those originating from the freedom of choosing
the representation of the enveloping algebra, already discussed
in~\cite{NCSM}.  The most serious aspect of such additional parameters
is the question to which extent deviations from SM predictions could
be hidden by particular choices of the ambiguous contributions, making
the NCSM untestable by experiment.

\subsection{Infinitesimal Gauge Transformations}
\label{sec:SWM-inf}

The non-linear gauge equivalence equations~(\ref{eq:SWM-condition})
are most easily solved by going over to the equivalent set of
equations for infinitesimal gauge transformations.  In the commutative
case, the latter are given by 
\begin{subequations}
\label{eq:delta}
\begin{align}
\label{eq:delta-A}
  \delta_\alpha A_\mu
     &= D_\mu^{\text{adj}}\alpha 
      = \partial_\mu\alpha - \ii\left[A_\mu,\alpha\right] \\
\label{eq:delta-psi}
  \delta_\alpha \psi &= \ii\alpha\psi\,,
\end{align}
\end{subequations}
while the noncommutative infinitesimal gauge transformations read
\begin{subequations}
\label{eq:hat-delta}
\begin{align}
\label{eq:hat-delta-A}
  \hat\delta_\alpha \hat A_\mu(A,\theta)
     &= \hat D_\mu^{\text{adj}}\hat\lambda(\alpha,A,\theta)
      = \partial_\mu\hat\lambda(\alpha,A,\theta)
         - \ii\left[\hat A_\mu(A,\theta) 
         \astcomma\hat\lambda(\alpha,A,\theta)\right] \\
\label{eq:hat-delta-psi}
  \hat\delta_\alpha \hat\psi(\psi,A,\theta)
     &= \ii\hat\lambda(\alpha,A,\theta)\ast\hat\psi(\psi,A,\theta)\,.
\end{align}
\end{subequations}
From~(\ref{eq:delta}) and~(\ref{eq:hat-delta}), 
one readily obtains the
infinitesimal versions of the gauge equivalence 
equations~(\ref{eq:SWM-condition-A})
and~(\ref{eq:SWM-condition-psi}):
\begin{subequations}
\label{eq:SWM-condition-infinitesimal}
\begin{align}
\label{eq:SWM-condition-infinitesimal-A}
  \hat\delta_\alpha \hat A_\mu(A,\theta)
    &= \delta_\alpha\hat A_\mu(A,\theta) \\
\label{eq:SWM-condition-infinitesimal-psi}
  \hat\delta_\alpha \hat \psi(\psi,A,\theta)
    &= \delta_\alpha\hat\psi(\psi,A,\theta)\,,
\end{align}
\end{subequations}
where the commutative gauge transformation~$\delta_\alpha$ acts on
the arguments of the noncommutative gauge and matter fields 
via the chain rule.

However, the equations~(\ref{eq:SWM-condition-infinitesimal}) are not
sufficient.  The existence of a noncommutative gauge
parameter~$\hat\lambda(\alpha,A,\theta)$ in~(\ref{eq:SWM-condition})
requires that the commutator of two infinitesimal gauge
transformations closes to another gauge
transformation just as in the commutative case:
\begin{equation}
\label{eq:closure}
  \left(   \hat\delta_{\alpha}\hat\delta_{\beta}
         - \hat\delta_{\beta}\hat\delta_{\alpha} \right) \hat\psi
  = \hat\delta_{-\ii[\alpha,\beta]} \hat\psi\,,
\end{equation}
where~$[\alpha,\beta]$ denotes the bracket in the commutative Lie algebra.
Applying~(\ref{eq:SWM-condition-infinitesimal-psi}) 
twice and taking into account that the commutative gauge
transformation of the noncommutative gauge
parameter~$\hat\lambda(\alpha,A,\theta)$ does not vanish because it
depends on~$A_\mu$, one has
\begin{multline}
     \hat\delta_{\alpha}\hat\delta_{\beta}\hat\psi
   = \ii \hat\delta_{\alpha} \left(\hat\lambda(\beta)\ast\hat\psi\right)
   = \ii \hat\delta_{\alpha}\hat\lambda(\beta)\ast\hat\psi
      + \ii \hat\lambda(\beta)\ast\hat\delta_{\alpha}\hat\psi\\
   = \ii \hat\delta_{\alpha}\hat\lambda(\beta)\ast\hat\psi
      -  \hat\lambda(\beta)\ast\hat\lambda(\alpha)\ast\hat\psi \,,
\end{multline}
where all unnecessary arguments have been omitted. 
Substituting the above in~(\ref{eq:closure}) 
and factoring out~$\hat\psi$, the
additional consistency condition reads
\begin{equation}
\label{eq:SWM-condition-infinitesimal-lambda}
    \delta_{\alpha}\hat\lambda(\beta,A,\theta)
  - \delta_{\beta}\hat\lambda(\alpha,A,\theta)
  - \ii \left[\hat\lambda(\alpha,A,\theta)
          \astcomma\hat\lambda(\beta,A,\theta)\right]
  = \hat\lambda(-\ii[\alpha,\beta],A,\theta)\,.
\end{equation}
The infinitesimal consistency
equations~(\ref{eq:SWM-condition-infinitesimal})
and~(\ref{eq:SWM-condition-infinitesimal-lambda})
still contain all orders of~$\theta$ and closed expressions
for their solutions are not yet known.   However, we can express the
SWM as formal power series in~$\theta$:
\begin{subequations}
\begin{align}
  \hat\lambda(\alpha,A,\theta) &= 
     \sum_{n=0}^\infty \lambda^{(n)}(\alpha,A,\theta)\\
  \hat A_\mu(A,\theta) &= \sum_{n=0}^\infty A_\mu^{(n)}(A,\theta)\\
  \hat\psi(\psi,A,\theta) &= \sum_{n=0}^\infty \psi^{(n)}(\psi,A,\theta)\,,
\end{align}
\end{subequations}
expand the
equations~(\ref{eq:SWM-condition-infinitesimal})
and~(\ref{eq:SWM-condition-infinitesimal-lambda})
and solve them order by order in~$\theta$, starting 
at $n=0$ with the commutative gauge theory: 
\begin{subequations}
\begin{align}
  \lambda^{(0)}(\alpha,A,\theta) &= \alpha \\
  A_\mu^{(0)}(A,\theta) &= A_\mu \\
  \psi^{(0)}(\psi,A,\theta) &= \psi\,.
\end{align}
\end{subequations}

\subsection{First Order in $\theta$}
\label{sec:SWM1}

Writing all terms involving the unknown
function~$\lambda^{(1)}$ on the left-hand side, the expansion
of~(\ref{eq:SWM-condition-infinitesimal-lambda}) results in the
inhomogeneous linear equation
\begin{multline}
\label{eq:SWM-condition-infinitesimal-lambda-1}
    \delta_{\alpha}\lambda^{(1)}(\beta,A,\theta)
  - \delta_{\beta}\lambda^{(1)}(\alpha,A,\theta)
  - \ii \left[\lambda^{(1)}(\alpha,A,\theta),\beta\right]
  - \ii \left[\alpha,\lambda^{(1)}(\beta,A,\theta)\right] \\
  - \lambda^{(1)}(-\ii[\alpha,\beta],A,\theta)
  = - \frac{\theta^{\mu\nu}}{2} 
         \left[\partial_\mu\alpha,\partial_\nu\beta\right]_+\,.
\end{multline}
The general solution is given by
\begin{equation}
\label{eq:SWM-lambda-1}
  \lambda^{(1)}(\alpha,A,\theta)
      = \frac{1}{4} \theta^{\mu\nu} [\partial_\mu \alpha, A_\nu]_+
      + \ii c_\lambda^{(1)} \theta^{\mu\nu} 
          \left[ \partial_\mu \alpha, A_\nu \right]
\end{equation}
involving one free parameter~$c_\lambda^{(1)}$.
Using this solution, we can proceed
analogously with~(\ref{eq:SWM-condition-infinitesimal}) and derive the
linear equations for~$A^{(1)}_\mu$,
\begin{subequations}
\begin{multline}
  \delta_\alpha A^{(1)}_\mu(A,\theta)
         + \ii\left[A_\mu^{(1)}(A,\theta),\alpha\right] \\
     = \partial_\mu \lambda^{(1)}
         - \ii\left[A_\mu,\lambda^{(1)}(\alpha,A,\theta)\right]
         + \frac{\theta^{\rho\sigma}}{2}
            \left[\partial_\rho A_\mu,\partial_\sigma\alpha\right]_+\,,
\end{multline}
and~$\psi^{(1)}$,
\begin{equation}
  \delta_\alpha \psi^{(1)}(\psi,A,\theta)
       - \ii\alpha\psi^{(1)}(\psi,A,\theta)
     =   \ii\lambda^{(1)}(\alpha,A,\theta)\psi
       - \frac{\theta^{\rho\sigma}}{2}
            \partial_\rho \alpha \partial_\sigma\psi\,,
\end{equation}
\end{subequations}
where the right-hand sides depend on the free
parameter~$c_\lambda^{(1)}$ via~$\lambda^{(1)}(\alpha,A,\theta)$.
For each value of~$c_\lambda^{(1)}$, we then find the general
solutions
\begin{subequations}
\begin{multline}
\label{eq:SWM-A-1}
  A^{(1)}_\rho(A,\theta) = 
          \frac{1}{4}\theta^{\mu\nu}[F_{\mu\rho},A_\nu]_+
          - \frac{1}{4}\theta^{\mu\nu}[A_\mu,\partial_\nu A_\rho]_+ \\
       + \ii c_\lambda^{(1)} \,\theta^{\mu\nu} 
                 [D_\rho^{\text{adj}} A_\mu, A_\nu]  
       - 2\ii c_A^{(1)}\,\theta^{\mu\nu} D_\rho^{\text{adj}} F_{\mu\nu}
\end{multline}
and
\begin{equation}
\label{eq:SWM-psi-1}
  \psi^{(1)}(\psi,A,\theta) = 
     - \frac{1}{2}\theta^{\mu\nu}A_\mu\partial_\nu\psi 
     + \frac{\ii}{8} \left( 1 - 4 c_\lambda^{(1)} \right)
           \theta^{\mu\nu} [A_\mu, A_\nu] \psi
     + \frac{c_\psi^{(1)}}{2}\theta^{\mu\nu}F_{\mu\nu}\psi\,,
\end{equation}
\end{subequations}
parametrized by the free parameters~$c_A^{(1)}$ and~$c_\psi^{(1)}$, 
respectively.  Notice
that all terms proportional to~$c_\lambda^{(1)}$, $c_A^{(1)}$,
or~$c_\psi^{(1)}$ are Lie algebra valued. Therefore, they correspond to
field reparametrizations and will cancel in observables.  In fact,
only the anticommutators in the expression~(\ref{eq:SWM-A-1})
for~$A^{(1)}_\rho(A,\theta)$ require the enveloping algebra and carry
the potential to affect observables.

\subsection{Second Order in $\theta$}
\label{sec:SWM2}

In second order in~$\theta$, we again start with the closure
relation~(\ref{eq:SWM-condition-infinitesimal-lambda})
for gauge transformations:
\begin{multline}
\label{eq:SWM-condition-infinitesimal-lambda-2}
        \delta_\alpha\lambda^{(2)}(\beta,A)
      - \delta_\beta\lambda^{(2)}(\alpha,A) \\
      - \ii \left[\alpha,\lambda^{(2)}(\beta,A)\right]
      - \ii \left[\lambda^{(2)}(\alpha,A),\beta\right]
      - \lambda^{(2)}(-i[\alpha,\beta],A) \\
  = - \frac{\ii}{8}\theta^{\mu\nu}\theta^{\kappa\lambda}
        \left[\partial_\mu\partial_\kappa\alpha,
         \partial_\nu\partial_\lambda\beta\right]
      + \ii \left[\lambda^{(1)}(\alpha,A),\lambda^{(1)}(\beta,A)\right] \\
      - \frac{1}{2}\theta^{\mu\nu}
           \left(\left[\partial_\mu\lambda^{(1)}
             (\alpha,A),\partial_\nu\beta\right]_+
               - \left[\partial_\mu\lambda^{(1)}
             (\beta,A),\partial_\nu\alpha\right]_+\right)\,,
\end{multline}
where we have suppressed the dependence on~$\theta$
in~$\lambda^{(k)}(\alpha,A)$ for brevity.  Again, the
right-hand side depends on the free
parameter~$c_\lambda^{(1)}$ via~$\lambda^{(1)}(\alpha,A)$.
For~$c_\lambda^{(1)}$ fixed, we find that the general hermitian
solution of~(\ref{eq:SWM-condition-infinitesimal-lambda-2})
involves 15~new free 
parameters,~$c_{\lambda,1}^{(2)}$, \ldots, $c_{\lambda,15}^{(2)}$.  
For the present discussion it is sufficient to
describe the general characteristics of the solutions and to give
explicit expressions only for 
a specific choice of the free parameters.
In appendix~\ref{app:SWM-lambda} we spell out the 
specific solution corresponding
to~$c_{\lambda,i}^{(2)}=0$ in the case~$c_\lambda^{(1)}=0$.  The
lengthy complete expression can be found in the appendix
of~\cite{Alboteanu:2007}.  It can be shown that the solutions given
in~\cite{Jurco:2001rq} and~\cite{Moller:2004qq} are both contained in
our 16-parameter family of solutions.

Proceeding with the expansion of the gauge equivalence
equations~(\ref{eq:SWM-condition-infinitesimal}) in powers of $\theta$, 
one obtains the second order equations for gauge and matter fields:
\begin{subequations}\label{eq:GE2}
\begin{multline}
\label{eq:GE2-A}
  \delta_\alpha A^{(2)}_\rho(A)- \ii[\alpha,A^{(2)}_\rho(A)]  \\
     = \partial_\rho\lambda^2(\alpha,A)
         - \ii [A_\rho, \lambda^{(2)}(\alpha,A)]
         - \ii [A^{(1)}_\rho(A), \lambda^{(1)}(\alpha,A)] \\
   + \frac{1}{2}\theta^{\mu\nu}
           \Big(\big[\partial_\mu A^{(1)}_\rho(A),\partial_\nu\alpha\big]_+
   +  \big[\partial_\mu A_\rho,\partial_\nu\lambda^{(1)}(\alpha,A)\big]_+\Big)
          + \frac{\ii}{8}\theta^{\mu\nu}\theta^{\kappa\lambda}
                  \partial_\mu\partial_\kappa A_\rho
                     \partial_\nu\partial_\lambda\alpha
\end{multline}
and
\begin{multline}
\label{eq:GE2-psi}
  \delta_\alpha\psi^{(2)}(A)-\ii\alpha\psi^{(2)}(A)
    =     \ii \lambda^{(1)}(\alpha,A)\psi^{(1)}(A)
        + \ii \lambda^2(\alpha,A)\psi \\
        - \frac{1}{2}\theta^{\mu\nu} 
            \big(\partial_\mu\lambda^{(1)}(\alpha,A)\partial_\nu\psi
        + \partial_\mu\alpha\partial_\nu\psi^{(1)}(A)\big)
        - \frac{\ii}{8}\theta^{\mu\nu}\theta^{\kappa\lambda}
             \partial_\mu\partial_\kappa \alpha 
               \partial_\nu\partial_\lambda \psi\,,
\end{multline}
\end{subequations}
where the right-hand sides now depend 
on~$c_\lambda^{(1)}$, $c_A^{(1)}$, $c_\psi^{(1)}$,
$c_{\lambda,1}^{(2)}$, \ldots, and~$c_{\lambda,15}^{(2)}$.

\subsubsection{Gauge Fields}
\label{sec:SWM2-A}

Substituting~(\ref{eq:SWM-lambda-1}) and~(\ref{eq:SWM-A-1})
for~$\lambda^{(1)}(\alpha,A)$ and 
$A_\mu^{(1)}(A)$, respectively, and the general 
solution~$\lambda^{(2)}(\alpha,A)$ from~\cite{Alboteanu:2007} 
in ~(\ref{eq:GE2-A}), one gets the general hermitian 
solution for $A_\mu^{(2)}(A)$ which depends on 6 additional 
parameters $c^{(2)}_{A,1,\pm}$. The latter define
a convenient basis for the solutions of the homogeneous equation: 
\begin{subequations}
\label{eq:c2/A}
\begin{align}
\label{eq:c2/A/1}
  A^{(2)}_{\rho,1,\pm} &=
     c^{(2)}_{A,1,\pm} \frac{\ii}{2} \theta^{\mu\nu} \theta^{\kappa\lambda}
     \Bigl[ D^{\text{adj}}_\kappa F_{\mu\nu}, F_{\lambda\rho} \Bigr]_\pm \\
  A^{(2)}_{\rho,2,\pm} &=
     c^{(2)}_{A,2,\pm} \frac{\ii}{2} \theta^{\mu\nu} \theta^{\kappa\lambda}
     \Bigl[ F_{\mu\kappa}, D^{\text{adj}}_\rho F_{\nu\lambda} \Bigr]_\pm\\
  A^{(2)}_{\rho,3,\pm} &=
     c^{(2)}_{A,3,\pm} \frac{\ii}{2} \theta^{\mu\nu} \theta^{\kappa\lambda}
     \Bigl[ F_{\kappa\lambda}, D^{\text{adj}}_\rho F_{\mu\nu} \Bigr]_\pm\,.
\end{align}
\end{subequations}
The three commutator terms~$A^{(2)}_{\rho,i,-}$ are contained in the
Lie algebra and correspond to field redefinitions that must cancel in
observables. In contrast, the three anticommutator
terms~$A^{(2)}_{\rho,i,+}$ need not be part of the Lie algebra and can,
in principle, give non-vanishing contributions to scattering
amplitudes as we will demonstrate in section~\ref{sec:example}.
An explicit expression for a specific solution of $A^{(2)}_{\mu}$
is given in appendix~\ref{app:SWM-A}.

The solution presented in~\cite{Jurco:2001rq} can be shown to be
contained in our set of general solutions~\cite{Alboteanu:2007}.  
In contrast, the solution presented in~\cite{Moller:2004qq} is not.
In fact, one can verify that the solution as written
in~\cite{Moller:2004qq} does not satisfy the gauge equivalence
equation~(\ref{eq:GE2-A}) and must therefore be considered incorrect.
In addition, the solutions~$A^{(2)}_{\rho,1,\pm}$ of the homogeneous
equation are missing. As we will see 
later,~$A^{(2)}_{\rho,1,+}$ plays a particularly important r\^ole.
While the size of the expressions makes it hard to pinpoint the actual
error in~\cite{Moller:2004qq}, we want to mention
that our results have been obtained and
verified by long but straightforward algebraic manipulations using
\texttt{FORM}~\cite{Vermaseren:FORM}.

\subsubsection{Matter Fields}
\label{sec:SWM2-psi}

Plugging~(\ref{eq:SWM-lambda-1}) for~$\lambda^{(1)}(\alpha,A)$,
~(\ref{eq:SWM-psi-1}) for~$\psi^1(\psi,A)$,
and the general solution for~$\lambda^{(2)}(\alpha,A)$
given in~\cite{Alboteanu:2007} 
into~(\ref{eq:GE2-psi}), we obtain a three-parameter
family of matter field SWM.  
A suitable basis for the solutions to the homogeneous equations is given by
\begin{subequations}
\begin{align}
  \psi^{(2)}_{1}  &=
     \ii c^{(2)}_{\psi,1}
      \theta^{\mu\nu}\theta^{\kappa\lambda} 
          (D^{\text{adj}}_\mu F_{\nu\kappa})D_\lambda\psi\\
  \psi^{(2)}_{2} &=
   -\frac{c^{(2)}_{\psi,2}}{4}
      \theta^{\mu\nu}\theta^{\kappa\lambda} F_{\mu\nu}F_{\kappa\lambda}\psi\\
  \psi^{(2)}_{3} &=
    \frac{c^{(2)}_{\psi,3}}{2}
      \theta^{\mu\nu}\theta^{\kappa\lambda} F_{\mu\kappa}F_{\nu\lambda}\psi\,.
\end{align}
\end{subequations}
This result confirms the corresponding result in~\cite{Moller:2004qq}.
In appendix~\ref{app:SWM-psi} we present a specific solution
for~$\psi^{(2)}$. For the general solution we again refer 
to~\cite{Alboteanu:2007}.  


\section{NCSM Lagrangian and Feynman Rules}
\label{sec:Feynman}

Following the prescription of~\cite{Seiberg:1999vs}
for constructing a NCQFT, we
replace all field products in the Yang-Mills Lagrangian
\begin{equation}
\label{eq:YM}
  L_{\text{YM}} = - \frac{1}{2} \tr \left( F_{\mu\nu} F^{\mu\nu} \right) 
                  + \bar\psi \left( \ii \fmslash{D} - m \right) \psi
\end{equation}
by $\ast$-products, and obtain an action for fields in the enveloping
algebra that is invariant under NC gauge transformations
and follows from the Lagrangian
\begin{equation}
  L_{\text{YM},\ast} =
     - \frac{1}{2} \tr \left( F_{\mu\nu,\ast} \ast F_\ast^{\mu\nu} \right) 
     + \bar\psi \ast \left( \ii \fmslash{D}  - m \right) \ast  \psi
\end{equation}
with~$F_{\mu\nu}=\ii[D_\mu,D_\nu]$
and~$F_{\mu\nu,\ast}=\ii[D_\mu\astcomma D_\nu]$, respectively.  In a
second step, we apply the SWM~(\ref{eq:SWM}) to obtain an action for
fields residing in the Lie algebra that is invariant under commutative gauge
transformations and results from
\begin{equation}
\label{eq:NCYM}
  L_{\text{NCYM}} =
     - \frac{1}{2} \tr \left( \hat F_{\mu\nu,\ast}(A) 
              \ast \hat F_\ast^{\mu\nu}(A) \right) 
     + \bar{\hat\psi}(\psi,A) \ast \left( \ii \hat{\fmslash{D}}(A)  - m \right)
            \ast \hat\psi(\psi,A)\,.
\end{equation}
From~(\ref{eq:NCYM}), one can derive the Feynman rules for the NC
extension of the original commutative gauge theory.  Since the terms
in~(\ref{eq:YM}) and~(\ref{eq:NCYM}) that are quadratic in the fields are
identical, we can perform gauge-fixing and introduce Faddeev-Popov
ghosts directly to~(\ref{eq:NCYM}) in terms of the commuting gauge
fields.  
Hence, the ghost interactions are not modified by NC contributions.

For the purposes of the present paper, we are mainly interested in the
cubic and quartic couplings in the neutral current sector of the NCSM
that contribute to boson pair production in fermion - antifermion
annihilation, $f \bar{f} \to V V$, at
tree level up to second order in~$\theta$.  With \emph{all} momenta
incoming, including outgoing fermions, we define the vertex factors as
follows:
\begin{subequations}
\label{eq:vertices}
\begin{align}
\label{eq:fVf}
  \parbox{28mm}{\includegraphics{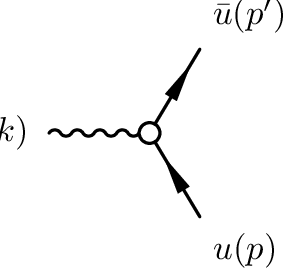}} &=
     \ii g \cdot V_{\mu}(p',k,p)\\
\label{eq:fVVf}
  \parbox{28mm}{\includegraphics{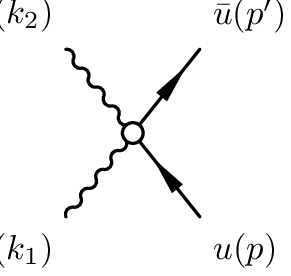}} &=
     \ii g^2 \cdot V_{\mu_2\mu_1}(p',k_2,k_1,p)\\
\label{eq:VVV}
  \parbox{28mm}{\includegraphics{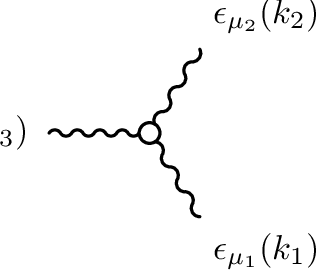}} &=
     \ii g_{[\rho]} \cdot V_{\mu_1\mu_2\mu_3}(k_1,k_2,k_3)\,.
\end{align}
\end{subequations}
where~$g_{[\rho]}$ denotes the three-gauge boson coupling that depends
on the choice for the representation~$\rho$ of the enveloping algebra.
In a~$U(1)$ gauge theory we can choose an arbitrary hermitian
matrix~$\rho(T)$ as a generator, normalized
to~$\tr\left(\rho(T)\rho(T)\right) = 1$.  Consequently the squares of
the eigenvalues of~$\rho(T)$ are bounded,~$0\le \lambda^2_i\le1$, and we
find $-g\le g_{[\rho]} \le g$, with~$g_{[\rho]}=0$
for~$\rho(T)=\sigma_3/\sqrt{2}$ and $g_{[\rho]}=g$ for~$\rho(T)=1$.
Only in the latter case, the anticommutator remains in the Lie algebra
representation.

Since the chiral structure of the fermionic currents remains
unaffected by the SWM, we have written the following vertex factors
for pure vector currents.  The necessary
substitutions $\gamma_\mu\to g_V\gamma_\mu - g_A\gamma_\mu\gamma_5$
depending on the fermion flavor and the type of vector boson coupled 
can be copied directly from the SM Lagrangian.
Using the notations~$p\theta q = p_\mu\theta^{\mu\nu}q_\nu$
and~$p\theta^\nu = p_\mu\theta^{\mu\nu}$, 
the vertex factors~(\ref{eq:vertices}) 
up to second order in $\theta$ are given by
\begin{subequations}
\begin{equation}
  V^{(1)}_{\mu}(p',k,p) =
      \frac{\ii}{2}
         \Bigl[
             k\theta_\mu \fmslash{p}(1 - 4 c_\psi^{(1)}) 
           + 2\,k\theta_\mu \fmslash{k}(c_A^{(1)} -  c_\psi^{(1)}) 
           - p\theta_\mu \fmslash{k}
           - (k\theta p) \gamma_{\mu}
        \Bigr]
\end{equation}
\begin{multline}
  V^{(2)}_{\mu}(p',k,p) = \\
     \frac{1}{8}(k\theta p)
         \Bigl[
             k\theta_\mu \fmslash{p}(1 - 16 c^{(2)}_\psi)
           + 4 k\theta_\mu \fmslash{k}(c_A^{(1)} - 2c^{(2)}_\psi)
           - p\theta_\mu \fmslash{k}
           - (k\theta p) \gamma_{\mu}
         \Bigr]
\end{multline}
\begin{multline}
 V^{(1)}_{\mu_2\mu_1}(p',k_2,k_1,p) = \\
      \frac{\ii}{2} \Bigl[
        k_2\theta_{\mu_1}\gamma_{\mu_2} 
               - k_1\theta_{\mu_1}\gamma_{\mu_2}(1 - 4 c_\psi^{(1)}) 
        -  \theta_{\mu_1\mu_2} \fmslash{k}_1   +
        (\mu_1\leftrightarrow\mu_2, k_1\leftrightarrow k_2)
             \Bigr]
\end{multline}
\begin{multline}
\label{eq:fbarVVf-on-shell}
 V^{(2)}_{\mu_2\mu_1}(p',k_2,k_1,p) =
         + \frac{1}{8}\Bigl[
              k_1\theta k_2\,k_1\theta_{\mu_1} \gamma_{\mu_2}\,
                 (8c^{(2)}_{A,1,+} - 4 c_\psi^{(1)} + 8c^{(2)}_\psi -1)\\
           +   k_1\theta p\,k_1\theta_{\mu_1} 
                          \gamma_{\mu_2}\,(16c^{(2)}_\psi - 1 )
       +   2\,k_2\theta p\,k_1\theta_{\mu_1} 
                          \gamma_{\mu_2}\,(4 c_\psi^{(1)} -1 )\\
                          - k_1\theta k_2 \,k_2\theta_{\mu_1}\gamma_{\mu_2}\,
                          + 3\,k_1\theta p \,k_2\theta_{\mu_1}\gamma_{\mu_2}\,
                          + 2\,k_2\theta p \,k_2\theta_{\mu_1}\gamma_{\mu_2}\
                          - 3\,k_1\theta k_2 \,p\theta_{\mu_1}\gamma_{\mu_2}\,
                     \\
        +  4\,k_1\theta_{\mu_1} k_1\theta_{\mu_2}\fmslash{k}_1
                  (2 c^{(2)}_{A,1,+} - c_A^{(1)} - c_\psi^{(1)})
        + 2\,k_1\theta_{\mu_1} p\theta_{\mu_2} 
                             \fmslash{k}_1(1 - 4 c_\psi^{(1)})\\
                   + 2\,k_2\theta_{\mu_1} p\theta_{\mu_2}\fmslash{k}_1 
                   - 4\,\theta_{\mu_1\mu_2} k_1\theta p\fmslash{k}_1
                   +({\mu_1}\leftrightarrow{\mu_2}, k_1\leftrightarrow k_2)
             \Bigr] \\
       + \text{terms vanishing by equations of motion}
\end{multline}
\begin{multline}
  V^{(1)}_{\mu_1\mu_2\mu_3}(k_1,k_2,k_3) = \\
        \theta_{\mu_1\mu_2}
          \left[ (k_1 k_3) k_{2,\mu_3} - (k_2 k_3) k_{1,\mu_3} \right]
      + \left( k_1 \theta k_2 \right)
          \left[ k_{3,\mu_1} g_{\mu_2\mu_3}
                   - g_{\mu_1\mu_3} k_{3,\mu_2} \right] \\
      + \Bigl[   (k_1\theta)_{\mu_1}
      \left[ k_{2,\mu_3} k_{3,\mu_2} - (k_2 k_3) g_{\mu_2\mu_3} \right]
               - (\mu_1 \leftrightarrow \mu_2)
               - (\mu_1 \leftrightarrow \mu_3)
        \Bigr] \\
  + \text{cyclical permutations of}\;
       \bigl\{ (\mu_1,k_1) , (\mu_2,k_2), (\mu_3,k_3) \, \bigr\}
\end{multline}
\begin{multline}
\label{eq:ggg2}
  V^{(2)}_{\mu_1\mu_2\mu_3}(k_1,k_2,k_3) = \\
      \ii \Bigl[
             k_1 \theta k_2  k_1 \theta_{\mu_1} \Big(
            (c^{(2)}_{A,1,+} - c^{(1)}_A)
            (k_{1,\mu_3} k_{3,\mu_2} - g_{\mu_2\mu_3} (k_1 k_3) )\\
          +  c^{(2)}_{A,1,+}(k_{2,\mu_3} k_{3,\mu_2} 
              - g_{\mu_2\mu_3} (k_2 k_3))
          \Big)\\ 
       +    k_1 \theta_{\mu_1}  k_1 \theta_{\mu_2}
            (c^{(2)}_{A,1,+} - c^{(1)}_A)
                            (k_{2,\mu_3} (k_1 k_3) - k_{1,\mu_3} (k_2 k_3) ) \\
       + \bigl((\mu_2,k_2)\leftrightarrow (\mu_3,k_3)\bigr) \Bigr] \\
  + \text{cyclical permutations of}\;
       \bigl\{ (\mu_1,k_1) , (\mu_2,k_2), (\mu_3,k_3) \, \bigr\}\,.
\end{multline}
\end{subequations}
Note that the triple gauge boson vertex~(\ref{eq:ggg2})
at~$\mathcal{O}(\theta^2)$ is generated by the SWM alone. 
There are no contributions from the Moyal-Weyl
$\ast$-product.
In this paper, we will apply the NCSM Feynman rules to the process $f
\bar{f} \to V V$ at tree level. In the above, we have therefore given
only the on-shell expression for the $\bar{\mathit{f}}\mathit{VVf}$ contact
term~(\ref{eq:fbarVVf-on-shell}), dropping terms which vanish by 
equation of motion.  The complete
expressions can be found in~\cite{Alboteanu:2007}.

\section{Influence of SWM Ambiguities on Observables}
\label{sec:ambiguities}

As noted in section~\ref{sec:SWM1}, all SWM ambiguities to first order 
in $\theta$, 
i.\,e.~all terms in the SWM proportional to~$c_\lambda^{(1)}$, $c_A^{(1)}$,
and~$c_\psi^{(1)}$ correspond to Lie algebra valued field
redefinitions and must cancel in observables such as on-shell scattering 
amplitudes to this order. This is explicitly checked
for~$f \bar{f} \to V V$ in~(\ref{eq:amplitude(1)}). Note, however, that the 
above~$\mathcal{O}(\theta)$ parameters reappear in the SWM in higher order.

\subsection{Ambiguities to second order in $\theta$}
\label{sec:ambiguities2}

\begin{figure}
  \begin{center}
    \hfil\\[\baselineskip]
    \includegraphics{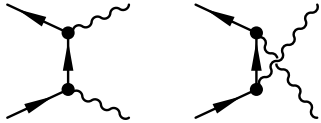}
  \end{center}
  \caption{\label{fig:diagrams0}Feynman diagrams contributing to
    $e^+e^-\to\gamma\gamma$ in~QED.}
\end{figure}

\begin{figure}
  \begin{center}
    \hfil\\[\baselineskip]
    \includegraphics{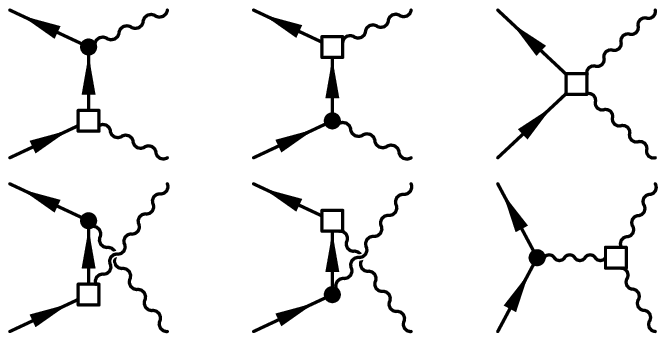}
  \end{center}
  \caption{\label{fig:diagrams1}Feynman diagrams contributing to
    $e^+e^-\to\gamma\gamma$ in~$\mathcal{O}(\theta)$ in~NCQED.  The open
    squares denote $\mathcal{O}(\theta)$-vertices.}
\end{figure}

\begin{figure}
  \begin{center}
    \hfil\\[\baselineskip]
    \includegraphics{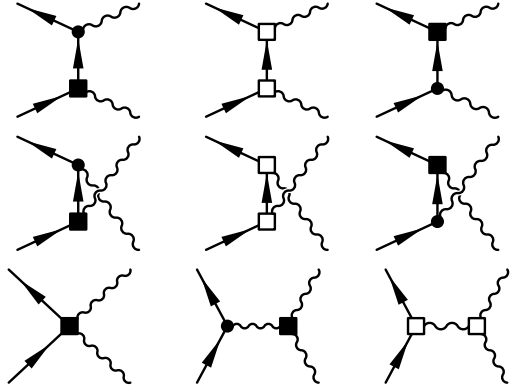}
  \end{center}
  \caption{\label{fig:diagrams2}Feynman diagrams contributing to
    $e^+e^-\to\gamma\gamma$ in $\mathcal{O}(\theta^2)$ in~NCQED. The filled
    squares denote $\mathcal{O}(\theta^2)$-vertices.}
\end{figure}

We have already pointed out, that beyond ~$\mathcal{O}(\theta)$ 
there are no such general arguments for or against the cancellation of
the SWM ambiguities that do not correspond to non-singular field 
redefinitions in observables. Therefore, we can approach this question 
presently only by studying specific examples. For that we choose the 
NCQED process $e^+e^-\to\gamma\gamma$ as a prototype process in the 
neutral current sector of the NCSM.
The additional $Z$-boson couplings, their chiral structure and the 
$Z$-mass in the NCSM will not add to our conclusions about the 
SWM ambiguities. The Feynman diagrams contributing
to~$e^+e^-\to\gamma\gamma$ in NCQED are depicted 
in~figs.~\ref{fig:diagrams0}-\ref{fig:diagrams2}. 
The Feynman rules to~$\mathcal{O}(\theta)$ and ~$\mathcal{O}(\theta^2)$
are given in section~\ref{sec:Feynman}.

Actually, it is possible to demonstrate the dependence of the
scattering amplitude on the free~$\mathcal{O}(\theta^2)$
parameter~$c^{(2)}_{A,1,+}$ 
and the interplay of repara\-metrization invariance and 
the enveloping algebra even without performing a complete 
calculation.  The term in~(\ref{eq:c2/A/1})
relevant for the tree level diagrams of~fig.~\ref{fig:diagrams2}
contains two gauge fields. Suppressing terms with more than two 
gauge fields, one has
\begin{equation}
\label{eq:A2-ambiguity}
  A^{(2)}_{\rho,1,+} =
    \ii c^{(2)}_{A,1,+} \theta^{\mu\nu}\theta^{\kappa\lambda}
    \partial_{\mu}\partial_{\kappa} A_{\nu}
  (\partial_{\lambda} A_{\rho} - \partial_{\rho} A_{\lambda}) + \ldots\,.
\end{equation}
This term contributes
both to the contact and the three-gauge boson vertex.  
Explicitly, from the
Feynman rule~(\ref{eq:fbarVVf-on-shell}) for the contact term 
one finds the following contribution to the scattering amplitude:
\begin{multline}
\label{eq:contact2r}
A^{(2)}_{\textrm{contact}} =
    \ii g^2  c^{(2)}_{A,1,+} \Bigl[
          k_1 \theta \varepsilon_1 (k_1 \theta k_2 \fmslash{\varepsilon}_2 
        + k_1 \theta \varepsilon_2 \fmslash{k}_1)  \\
        + k_2 \theta \varepsilon_2 (k_2 \theta k_1 \fmslash{\varepsilon}_1 
        + k_2 \theta \varepsilon_1 \fmslash{k}_2) \Bigr] + \ldots\,.
\end{multline}
Since~$c^{(2)}_{A,1,+}$ is absent from the on-shell
$\bar{f}\mathit{Vf}$-vertex, this contribution cannot be
cancelled by any term coming from the $t$- or $u$-channel
diagrams. Therefore, a cancellation must involve $s$-channel diagrams,
which however are proportional to the representation dependent 
coupling~$g_{[\rho]}$ as can be seen from~(\ref{eq:VVV}).
Consequently, the cancellation 
can at most take place for a particular value of~$g_{[\rho]}$,
namely for~$g_{[\rho]}=g$,
when the noncommutative gauge fields do not leave the Lie algebra 
and the SWM are just field reparametrizations.

As a side remark, the non-vanishing, a priori undetermined 
contribution~(\ref{eq:contact2r}) 
to the tree level amplitude of~$f\bar f\to \mathrm{V}\mathrm{V}'$ 
results from one of the SWM ambiguities missed in~\cite{Moller:2004qq}. 

\subsection{$e^+e^-\to\gamma\gamma$ Scattering Amplitude}
\label{sec:example}

We will now corroborate these preliminary remarks by a complete
tree level calculation of~$e^+e^-\to\gamma\gamma$
up to second order in $\theta$. 
It is useful to split the full scattering amplitude in the
following pieces: 
\begin{equation}
  A(e^+e^-\to\gamma\gamma) =
     g^2 A^{\textrm{SM}} + g^2 A^{(1)} + gg_{[\rho]} A_s^{(1)}
   + g^2 A^{(2)} + gg_{[\rho]} A_s^{(2)}\,,
\end{equation}
which are self-explaining.
It should be noted that the $s$-channel contributions~$A_s^{(i)}$
must be separately gauge invariant, because their normalization
depends on the choice of the representation of the enveloping algebra.

For completeness, we restate the familiar QED amplitude from the
diagrams in fig.~\ref{fig:diagrams0}:
\begin{equation}
  A^{\textrm{SM}} =
    - \frac{\ii}{q_u^2} \bar{v}(p_2)
         \fmslash{\varepsilon}_1\fmslash{q}_u\fmslash{\varepsilon}_2 u(p_1)
    - \frac{\ii}{q_t^2} \bar{v}(p_2)
         \fmslash{\varepsilon}_2\fmslash{q}_t\fmslash{\varepsilon}_1 u(p_1)\,,
\end{equation}
with~$q_t=p_1-k_1$ and~$q_u=p_1-k_2$, as well as the
$\mathcal{O}(\theta)$-amplitude in NCQED~\cite{Ohl/Reuter:2004:NCPC}:
\begin{multline}
\label{eq:amplitude(1)}
  A^{(1)} =
   - \frac{\ii}{q_u^2} \left(\frac{p_1 \theta p_2 
     + k_1 \theta k_2}{2\ii}\right)
       \bar{v}(p_2) \fmslash{\varepsilon}_1
                 \fmslash{q}_u\fmslash{\varepsilon}_2 u(p_1) \\
   - \frac{\ii}{q_t^2} \left(\frac{p_1 \theta p_2 
              - k_1 \theta k_2}{2\ii}\right)
       \bar{v}(p_2) \fmslash{\varepsilon}_2
              \fmslash{q}_t\fmslash{\varepsilon}_1 u(p_1)
   + A_{\text{no pole}}^{(1)}\,,
\end{multline}
with
\begin{equation}
\label{eq:A-nopole}
   A_{\text{no pole}}^{(1)} =
     \bar{v}(p_2) \left[
        \frac{1}{2}\varepsilon_1 \theta 
                         \varepsilon_2 (\fmslash{k}_1-\fmslash{k}_2)
        - k_1 \theta \varepsilon_2 \fmslash{\varepsilon}_1
        - k_2 \theta \varepsilon_1 \fmslash{\varepsilon}_2 \right] u(p_1)\,,
\end{equation}
which collects all contributions  
from the $t$- and $u$-channel diagrams as well as the contact diagram 
in fig.~\ref{fig:diagrams1} containing no pole.
As expected, all contributions from ambiguous terms in the SWM 
cancel in~$A^{(1)}$ after application of the equations of motion.  
Note that the $1/t$- and $1/u$-pole terms in~(\ref{eq:amplitude(1)})
result solely from the expansion of the Moyal phases to 
order $\theta$.  
This can be easily seen by combining the phase factors
which multiply the vertices in a given diagram. For
momenta~$p_1$, $p_2$ and~$p_3$ flowing into a vertex 
momentum conservation implies the identities
\begin{equation}
  \ee^{-\ii p_1\theta p_2} = \ee^{-\ii p_2\theta p_3} 
   = \ee^{-\ii p_3\theta p_1}\,.
\end{equation}
In the $t$-channel
one thus obtains the total phase factor 
\begin{equation}
  \ee^{-\ii (-k_2)\theta(p_1-k_1)} \ee^{-\ii p_1\theta(p_2-k_2)}
    = \ee^{-\ii (p_1\theta p_2 - k_1\theta k_2)}\,,
\end{equation}
from which the corresponding phase factor in the $u$-channel follows 
by exchanging~$k_1\leftrightarrow k_2$:
\begin{equation}
   \ee^{-\ii (p_1\theta p_2 + k_1\theta k_2)}\,.
\end{equation}

Turning to the $s$-channel amplitude $A_s^{(1)}$, one finds
that the $1/s$-pole contributions cancel in all contributions from the
SWM yielding
\begin{equation}
\label{eq:amplitude(1)-s-channel}
  A_s^{(1)} = A_{s,\ast}^{(1)} - A_{\text{no pole}}^{(1)}\,,
\end{equation}
with the $1/s$-pole coming from the Moyal-Weyl $\ast$-product alone:
\begin{equation}
\label{eq:amplitude(1)-s-channel-*}
  A_{s,\ast}^{(1)} =
    \frac{1}{s} (k_1\theta k_2)\, \bar{v}(p_2)
       \left[ \frac{1}{2}
         (\varepsilon_1\varepsilon_2)(\fmslash{k}_2-\fmslash{k}_1) 
               + (k_1\varepsilon_2) \fmslash{\varepsilon}_1
               - (k_2\varepsilon_1) \fmslash{\varepsilon}_2 \right] u(p_1) \,,
\end{equation}
and $A_{\text{no pole}}^{(1)}$ given in~(\ref{eq:A-nopole}).
This ensures that all effects from the SWM cancel 
for~$g_{[\rho]}=g$ as they should, since in this case the
SWM are non-singular field reparametrizations.

To second order in $\theta$, we obtain from the 
diagrams in fig.~\ref{fig:diagrams2}
\begin{multline}
\label{eq:amplitude(2)}
   A^{(2)} =
     - \frac{\ii}{q_u^2} \frac{1}{2}
       \left(\frac{p_1 \theta p_2 + k_1 \theta k_2}{2\ii}\right)^2
        \bar{v}(p_2)\fmslash{\varepsilon}_1
         \fmslash{q}_u\fmslash{\varepsilon}_2 u(p_1)\\
     - \frac{\ii}{q_t^2} \frac{1}{2}
       \left(\frac{p_1 \theta p_2 - k_1 \theta k_2}{2\ii}\right)^2
        \bar{v}(p_2) \fmslash{\varepsilon}_2\fmslash{q}_t
         \fmslash{\varepsilon}_1 u(p_1)
     + A_{\text{no pole}}^{(2)}
\end{multline}
with
\begin{multline}
\label{eq:A2-nopole}
   A_{\text{no pole}}^{(2)} =
    - \frac{\ii}{2} p_1 \theta p_2  A_{\text{no pole}}^{(1)}
     +\ii(c^{(2)}_{A,1,+} - c_A^{(1)}) \bar{v}(p_2)
        \bigg[
            k_1 \theta \varepsilon_1 k_1 \theta \varepsilon_2 \fmslash{k}_1 
          + k_2 \theta \varepsilon_1 k_2 \theta \varepsilon_2 \fmslash{k}_2 \\
          + k_1 \theta k_2
              \Bigl(   k_1 \theta \varepsilon_1 \fmslash{\varepsilon}_2 
                     - k_2 \theta \varepsilon_2 \fmslash{\varepsilon}_1 \Bigr)
        \bigg] u(p_1)\,.
\end{multline}
Note that the $1/t$- and $1/u$-pole terms in~(\ref{eq:amplitude(2)})
follow again from the expansion of the combined Moyal phases 
and contain no contribution from the SWM.  
However, in contrast to~$A_{\text{no pole}}^{(1)}$, 
the second order amplitude~$A_{\text{no pole}}^{(2)}$ 
does depend on ambiguous terms in the SWM. In the case at hand,
this is signalled by the appearance of the free 
parameters~$c_A^{(1)}$ and~$c^{(2)}_{A,1,+}$.  The exact cancellation 
for the choice~$c^{(2)}_{A,1,+}= c_A^{(1)}$ appears
to be accidental.

As in first order in $\theta$, we find that all $1/s$-poles cancel in the
$s$-channel terms coming from the SWM.
Since the Moyal-Weyl $\ast$-product only contributes
to the three-photon couplings in odd orders of $\theta$
there is no $1/s$-pole term at all in $\mathcal{O}(\theta^2)$.
The result is exactly the negative of~\ref{eq:A2-nopole}: 
\begin{equation}
   A_s^{(2)} = - A_{\text{no pole}}^{(2)}\,,
\end{equation}
leading again to the cancellation of all
contributions from the SWM in the case~$g_{[\rho]}=g$,
including the ambiguous terms involving the free  
parameter~$c^{(2)}_{A,1,+} - c_A^{(1)}$.  
However, in general, the cross section for~$e^+e^-\to\gamma\gamma$ 
is affected by this SWM ambiguity.

\section{Conclusions}
\label{sec:summary}

We have investigated the noncommutative extension of the standard model 
up to second order in $\theta$.  As our main result, we find that the
general solution for the corresponding Seiberg-Witten maps contains 
more ambiguous terms than those reported previously, and that the SWM
ambiguities do not necessarily cancel in observables.
Furthermore, studying the scattering amplitude for~$e^+e^-\to\gamma\gamma$ 
as an explicit example we have shown that the ambiguities
remaining in the scattering amplitude 
can be traced to the necessary extension of the Lie
algebra of the gauge group to its enveloping algebra, which elevates
the SWM from a unobservable field reparametrization to a source of
new effective interactions.

Our results imply that the parameter space of 
the NCSM~\cite{NCSM} in~$\mathcal{O}(\theta^2)$ 
is larger than assumed so far.  There is every
reason to expect that higher orders in~$\theta$ will introduce even
more ambiguities and evidence for this has been found
in~NCQED~\cite{Ohl/Rauh/Rueckl/Zeiner:2007:AOSWM}.

As an outlook, phenomenological consequences of the NCSM
at~$\mathcal{O}(\theta^2)$ with the focus on collider searches 
will be presented in an upcoming 
paper~\cite{Alboteanu/Ohl/Rueckl:2007:Pheno}.

\subsection*{Acknowledgments}
This research is supported by Deutsche Forschungsgemeinschaft,
grant RU 311/1-1 and Research Training Group 1147 \textit{Theoretical
Astrophysics and Particle Physics}, and by Bundesministerium f\"ur 
Bildung und Forschung
Germany, grant 05H4\-WWA/2.  A.\,A.~gratefully
acknowledges support from Evangelisches Studienwerk e.\,V.~Villigst.
R.R. thanks the members of SLAC Theory Group for their kind hospitality.

\appendix
\section{Seiberg-Witten Maps in~$\mathcal{O}(\theta^2)$}
\label{app:SWM}

\subsection{Gauge Parameter}
\label{app:SWM-lambda}

For each value of~$c_\lambda^{(1)}$, we find a family of
hermitian solutions
to~(\ref{eq:SWM-condition-infinitesimal-lambda-2}) depending on
the free parameters~$c_{\lambda,1}^{(2)}$, \ldots,
$c_{\lambda,15}^{(2)}$. The specific solution corresponding
to~$c_{\lambda,i}^{(2)}=0$ for the case~$c_\lambda^{(1)}=0$
is given by
\begin{multline}
\label{eq:lambda2-special}
  \lambda^{(2)}(\alpha,A) =
     \frac{\ii}{32} \theta^{\kappa\lambda} \theta^{\mu\nu} \Big(
          -  3 A_\kappa  A_\lambda  \partial_\nu \alpha A_\mu
          -  4 A_\kappa  A_\nu  \partial_\lambda \alpha A_\mu
          -  3 A_\kappa  \partial_\lambda \alpha A_\mu  A_\nu\\
          -  2 A_\lambda  \partial_\nu \alpha A_\mu  A_\kappa
          -  2 A_\mu  A_\kappa  A_\lambda  \partial_\nu \alpha
          -    A_\mu  A_\nu  A_\kappa  \partial_\lambda \alpha \\
          -  2 A_\nu  A_\kappa  \partial_\lambda \alpha A_\mu 
          -  4 A_\nu  \partial_\lambda \alpha A_\mu  A_\kappa
          -  2 \partial_\mu A_\kappa \partial_\lambda \partial_\nu \alpha \\
          -  2 \partial_\lambda \alpha A_\mu  A_\nu  A_\kappa
          -    \partial_\nu \alpha A_\mu  A_\kappa  A_\lambda
          +  2 \partial_\lambda \partial_\nu \alpha  \partial_\mu A_\kappa
          \Big)\\
   + \frac{1}{16} \theta^{\kappa\lambda} \theta^{\mu\nu}  \Big(
            4 A_\kappa  \partial_\lambda \partial_\nu \alpha  A_\mu
          +   A_\lambda  \partial_\nu \alpha \partial_\mu A_\kappa
          + 2 A_\mu  A_\kappa  \partial_\lambda \partial_\nu \alpha \\
          - 2 A_\mu  \partial_\kappa A_\nu \partial_\lambda \alpha
          -   \partial_\kappa A_\nu \partial_\lambda \alpha A_\mu
          +   \partial_\mu A_\kappa A_\lambda  \partial_\nu \alpha \\
          -   \partial_\lambda \alpha A_\mu  \partial_\kappa A_\nu
          + 2 \partial_\nu \alpha \partial_\mu A_\kappa  A_\lambda
          + 2 \partial_\lambda \partial_\nu \alpha  A_\mu  A_\kappa
          \Big)\,.
\end{multline}
For the general expression we refer to the appendix of~\cite{Alboteanu:2007}.

\subsection{Gauge Fields}
\label{app:SWM-A}
A representative of the six-parameter family of second order
SWM~$A^{(2)}_\rho(A)$ corresponding to the parameter choice
$c_\lambda^{(1)}=c_{\lambda,i}^{(2)}=c_A^{(1)}=0$ is given by
\begin{align}
\label{eq:A2-special}
  A^{(2)}_\rho(A) = 
        \frac{\ii}{16} \theta^{\mu\nu} \theta^{\kappa\lambda}  & \Big(
            2 [\partial_\nu \partial_\lambda A_\rho, \partial_\mu A_\kappa]  
          +  [\partial_\mu A_\kappa, \partial_\rho \partial_\lambda A_\nu]  \Big)\\ \nonumber
       + \frac{1}{16}\theta^{\mu\nu}\theta^{\kappa\lambda}   &\Big(
          +  2A_\mu A_\kappa \partial_\nu \partial_\lambda A_\rho
          -  A_\mu A_\kappa \partial_\nu \partial_\rho A_\lambda
          +  A_\mu A_\kappa \partial_\rho \partial_\lambda A_\nu\\ \nonumber&
          +  4A_\mu \partial_\nu A_\kappa \partial_\lambda A_\rho
          -  4A_\mu \partial_\nu A_\kappa \partial_\rho A_\lambda
          -  2A_\mu \partial_\kappa A_\nu \partial_\lambda A_\rho\\ \nonumber &
          -  2A_\nu \partial_\lambda A_\rho \partial_\mu A_\kappa
          +  3A_\nu \partial_\rho A_\lambda \partial_\mu A_\kappa
          +  4A_\kappa \partial_\nu \partial_\lambda A_\rho A_\mu\\ \nonumber &
          +  2A_\lambda \partial_\nu A_\rho \partial_\mu A_\kappa
          -  A_\lambda \partial_\rho A_\nu \partial_\mu A_\kappa
          -  A_\rho \partial_\mu A_\kappa \partial_\lambda A_\nu\\ \nonumber &
          -  4\partial_\mu A_\kappa A_\nu \partial_\lambda A_\rho
          +  \partial_\mu A_\kappa A_\nu \partial_\rho A_\lambda
          +  \partial_\mu A_\kappa A_\lambda \partial_\rho A_\nu\\ \nonumber &
          -  \partial_\mu A_\kappa \partial_\lambda A_\nu A_\rho
          +  2\partial_\nu A_\kappa \partial_\lambda A_\rho A_\mu
          -  3\partial_\nu A_\kappa \partial_\rho A_\lambda A_\mu\\ \nonumber &
          +  2\partial_\nu A_\rho \partial_\mu A_\kappa A_\lambda
          -  2\partial_\kappa A_\nu \partial_\lambda A_\rho A_\mu
          +  \partial_\kappa A_\nu \partial_\rho A_\lambda A_\mu\\ \nonumber &
          +  2\partial_\lambda A_\nu A_\rho \partial_\mu A_\kappa
          +  2\partial_\lambda A_\rho A_\mu \partial_\nu A_\kappa
          -  4\partial_\lambda A_\rho \partial_\mu A_\kappa A_\nu\\ \nonumber &
          -  \partial_\rho A_\lambda A_\mu \partial_\nu A_\kappa
          -  \partial_\rho A_\lambda A_\mu \partial_\kappa A_\nu
          +  4\partial_\rho A_\lambda \partial_\mu A_\kappa A_\nu\\ \nonumber &
          +  2\partial_\nu \partial_\lambda A_\rho A_\mu A_\kappa
          +  \partial_\nu \partial_\rho A_\lambda A_\mu A_\kappa
          -  \partial_\rho \partial_\lambda A_\nu A_\mu A_\kappa \Big)\\ 
                                                                    \nonumber
       + \frac{\ii}{32} \theta^{\mu\nu}\theta^{\kappa\lambda} &  \Big(
          -  4A_\mu A_\nu A_\kappa \partial_\lambda A_\rho  
          +  3A_\mu A_\nu A_\kappa \partial_\rho A_\lambda  
      -  2\partial_\rho A_\lambda A_\mu A_\kappa A_\nu  \\ \nonumber &
          +  4A_\mu A_\kappa A_\nu \partial_\lambda A_\rho  
          -  2A_\mu A_\kappa A_\nu \partial_\rho A_\lambda  
          -  4A_\mu A_\kappa A_\lambda \partial_\nu A_\rho  \\ \nonumber &
          -  2A_\mu A_\kappa \partial_\nu A_\lambda A_\rho  
          +  2A_\mu A_\kappa \partial_\lambda A_\nu A_\rho  
          -  8A_\mu \partial_\nu A_\kappa A_\lambda A_\rho  \\ \nonumber &
          -  4A_\nu A_\kappa \partial_\lambda A_\rho A_\mu  
          +  4A_\nu A_\kappa \partial_\rho A_\lambda A_\mu  
          +  8A_\nu A_\lambda A_\rho \partial_\mu A_\kappa  \\ \nonumber &
          -  4A_\nu \partial_\lambda A_\rho A_\mu A_\kappa  
          -  2A_\nu \partial_\rho A_\lambda A_\mu A_\kappa  
          -  4A_\kappa A_\nu \partial_\lambda A_\rho A_\mu  \\ \nonumber &
          -  2A_\kappa A_\nu \partial_\rho A_\lambda A_\mu  
          -  4A_\kappa A_\lambda \partial_\nu A_\rho A_\mu  
          +  A_\kappa A_\lambda \partial_\rho A_\nu A_\mu  \\ \nonumber &
          -  8A_\kappa \partial_\lambda A_\nu A_\rho A_\mu  
          -  4A_\kappa \partial_\lambda A_\rho A_\mu A_\nu  
          +  A_\kappa \partial_\rho A_\lambda A_\mu A_\nu  \\ \nonumber &
          -  4A_\lambda A_\nu A_\rho \partial_\mu A_\kappa  
          -  4A_\lambda A_\rho A_\mu \partial_\nu A_\kappa  
          +  4A_\lambda A_\rho A_\mu \partial_\kappa A_\nu  \\ \nonumber &
          +  8A_\lambda A_\rho \partial_\mu A_\kappa A_\nu  
          -  4A_\lambda \partial_\nu A_\rho A_\mu A_\kappa  
          +  4A_\lambda \partial_\rho A_\nu A_\mu A_\kappa  \\ \nonumber &
          -  2A_\rho A_\mu A_\kappa \partial_\nu A_\lambda  
          -  2A_\rho A_\mu A_\kappa \partial_\lambda A_\nu  
          +  8A_\rho A_\mu \partial_\kappa A_\nu A_\lambda  \\ \nonumber &
          -  2A_\rho \partial_\mu A_\kappa A_\nu A_\lambda  
          +  2A_\rho \partial_\mu A_\kappa A_\lambda A_\nu  
          +  2\partial_\mu A_\kappa A_\nu A_\lambda A_\rho  \\ \nonumber &
          +  2\partial_\mu A_\kappa A_\lambda A_\nu A_\rho  
          -  4\partial_\nu A_\kappa A_\lambda A_\rho A_\mu  
          +  4\partial_\nu A_\lambda A_\rho A_\mu A_\kappa  \\ \nonumber &
          -  4\partial_\nu A_\rho A_\mu A_\kappa A_\lambda  
          +  4\partial_\kappa A_\nu A_\lambda A_\rho A_\mu  
          -  8\partial_\lambda A_\nu A_\rho A_\mu A_\kappa  \\ \nonumber &
          -  4\partial_\lambda A_\rho A_\mu A_\nu A_\kappa  
          +  4\partial_\lambda A_\rho A_\mu A_\kappa A_\nu  
          +  3\partial_\rho A_\nu A_\mu A_\kappa A_\lambda  \Big)\\ \nonumber
       + \frac{1}{32}\theta^{\mu\nu}\theta^{\kappa\lambda}  & \Big(
          -  3A_\mu A_\nu A_\kappa A_\lambda A_\rho
          +  2A_\mu A_\kappa A_\nu A_\lambda A_\rho
          -  4A_\nu A_\kappa A_\lambda A_\rho A_\mu\\ \nonumber &
          +  4A_\nu A_\lambda A_\rho A_\mu A_\kappa
          -  4A_\nu A_\rho A_\mu A_\kappa A_\lambda
          +  4A_\kappa A_\nu A_\lambda A_\rho A_\mu\\ \nonumber &
          -  4A_\kappa A_\lambda A_\nu A_\rho A_\mu
          -  2A_\kappa A_\lambda A_\rho A_\mu A_\nu
          -  8A_\lambda A_\nu A_\rho A_\mu A_\kappa\\ \nonumber &
          -  4A_\lambda A_\rho A_\mu A_\nu A_\kappa
          +  4A_\lambda A_\rho A_\mu A_\kappa A_\nu\\ \nonumber &
          -  3A_\rho A_\mu A_\nu A_\kappa A_\lambda 
          +  2A_\rho A_\mu A_\kappa A_\nu A_\lambda\Big)\,.
\end{align}
Here, we have arranged the terms according to the power of~$A_\mu$,
for the benefit of deriving Feynman rules.  
The general solution can be found in~\cite{Alboteanu:2007}.
As pointed out in
section~\ref{sec:SWM2-A}, this solution is related to the one
presented in~\cite{Jurco:2001rq}, but incompatible with the one
in~\cite{Moller:2004qq}.

\subsection{Matter Fields}
\label{app:SWM-psi}

Choosing~$c_\lambda^{(1)}=c_{\lambda,i}^{(2)}=c_\psi^{(1)}=0$, we obtain
the following representative of the three-parameter family of matter 
field SWM:
\begin{align}
\label{eq:psi2-special}
  \psi^{(2)}(\psi,A) = 
       \frac{\ii}{8} \theta^{\mu\nu}\theta^{\kappa\lambda} & \Big(
          -  \partial_\mu A_\kappa \partial_\nu 
                       \partial_\lambda \psi \Big) \\ \nonumber 
      +\frac{1}{16} \theta^{\mu\nu}\theta^{\kappa\lambda} & \Big(
          -  2A_\mu \partial_\kappa A_\nu \partial_\lambda\psi
          -  2\partial_\mu A_\kappa A_\nu \partial_\lambda\psi \\ \nonumber &
          +  2A_\mu A_\kappa \partial_\nu \partial_\lambda \psi
          +  4A_\mu \partial_\nu A_\kappa \partial_\lambda\psi
          -  \partial_\mu A_\kappa \partial_\nu A_\lambda\psi\Big) \\ 
                                                                   \nonumber 
      +\frac{\ii}{8} \theta^{\mu\nu}\theta^{\kappa\lambda} & \Big(
          -  2A_\mu \partial_\nu A_\kappa A_\lambda \psi 
          +  \partial_\mu A_\kappa A_\nu A_\lambda \psi\\ \nonumber & 
          -  A_\mu A_\kappa A_\lambda \partial_\nu\psi 
          +  A_\mu A_\kappa A_\nu \partial_\lambda\psi 
          -  A_\mu A_\nu A_\kappa \partial_\lambda\psi \Big) \\ \nonumber 
      +\frac{1}{32} \theta^{\mu\nu}\theta^{\kappa\lambda} & \Big(
          -  3A_\mu A_\nu A_\kappa A_\lambda \psi
          +  4A_\mu A_\kappa A_\nu A_\lambda \psi
          -  2A_\mu A_\kappa A_\lambda A_\nu \psi
      \Big)\,.
\end{align}
Again, the terms are ordered according to the power in the gauge field.
For the general solution one may consult~\cite{Alboteanu:2007}.



\begin{thebibliography}{99}

\bibitem{Seiberg:1999vs}
  N.~Seiberg and E.~ Witten,
  JHEP \textbf{09}, 32 (1999)
  [arXiv:hep-th/9908142].

\bibitem{Snyder}
  H.~S.~Snyder,
  Phys.{} Rev.{} \textbf{71}, 38 (1947).

\bibitem{Jurco:2001rq}
  B.~Jurco, L.~M\"oller, S.~Schraml, P.~Schupp and J.~Wess,
  Eur.{} Phys.{} J.{}  C \textbf{21}, 383 (2001)
  [arXiv:hep-th/0104153].

\bibitem{NCSM}
  X.~Calmet, B.~Jur{\v c}o, P.~Schupp, J.~Wess and M.~Wohlgenannt,
  Eur.{} Phys.{} J.{} \textbf{C23}, 363 (2002) 
  [arXiv:hep-ph/0111115];
  B.~Melic, K.~Passek-Kumericki, J.~Trampetic, P.~Schupp and M.~Wohlgenannt,
  Eur.{} Phys.{} J.{}  C \textbf{42}, 483 (2005)
  [arXiv:hep-ph/0502249];
  B.~Melic, K.~Passek-Kumericki, J.~Trampetic, P.~Schupp and M.~Wohlgenannt,
  Eur.{} Phys.{} J.{}  C \textbf{42}, 499 (2005)
  [arXiv:hep-ph/0503064].

\bibitem{Anomaly-Freedom}
  C.~P.~Martin,
  Nucl.{} Phys.{}  \textbf{B652}, 72 (2003)
  [arXiv:hep-th/0211164];
  F.~Brandt, C.~P.~Martin and F.~R.~Ruiz,
  JHEP \textbf{0307}, 068 (2003)
  [arXiv:hep-th/0307292].

\bibitem{Chaichian:2001py}
  M.~Chaichian, P.~Presnajder, M.~M.~Sheikh-Jabbari and A.~Tureanu,
  Eur.{} Phys.{} J.{} C \textbf{29}, 413 (2003)
  [arXiv:hep-th/0107055].

\bibitem{NCSM-renormalization:bosons}
  M.~Buric, D.~Latas and V.~Radovanovic,
  JHEP \textbf{0602}, 046 (2006)
  [arXiv:hep-th/0510133];
  M.~Buric, V.~Radovanovic and J.~Trampetic,
  JHEP \textbf{0703}, 030 (2007)
  [arXiv:hep-th/0609073];
  D.~Latas, V.~Radovanovic and J.~Trampetic,
  [arXiv:hep-th/0703018].

\bibitem{NCSM-renormalization:fermions}
  R.~Wulkenhaar,
  JHEP \textbf{0203}, 024 (2002)
  [arXiv:hep-th/0112248].

\bibitem{Renormalizability}
  H.~Grosse and R.~Wulkenhaar,
  Lett.{} Math.{} Phys.{} \textbf{71}, 13 (2005);
  J.{} Nonlin.{} Math.{} Phys.{}  \textbf{11} Suppl., 9 (2004);
  Commun.{} Math.{} Phys.{} \textbf{256}, 305 (2005)
  [arXiv:hep-th/0401128];
  V.~Rivasseau, F.~Vignes-Tourneret and R.~Wulkenhaar,
  Commun.{} Math.{} Phys.{} \textbf{262}, 565 (2006)
  [arXiv:hep-th/0501036];
  H.~Grosse and M.~Wohlgenannt,
  [arXiv:0706.2167].

\bibitem{NCSM-Pheno}
  W.~Behr, N.~G.~Deshpande, G.~Duplancic, P.~Schupp, J.~Trampetic and J.~Wess,
  Eur.{} Phys.{} J.{}  C \textbf{29}, 441 (2003)
  [arXiv:hep-ph/0202121];
  G.~Duplancic, P.~Schupp and J.~Trampetic,
  Eur.{} Phys.{} J.{}  C \textbf{32}, 141 (2003) 
  [arXiv:hep-ph/0309138];
  P.~Schupp, J.~Trampetic, J.~Wess and G.~Raffelt,
  Eur.{} Phys.{} J.{}  C \textbf{36}, 405 (2004)
  [arXiv:hep-ph/0212292];
  P.~Minkowski, P.~Schupp and J.~Trampetic,
  Eur.{} Phys.{} J.{}  C \textbf{37}, 123 (2004) 
  [arXiv:hep-th/0302175];
  B.~Melic, K.~Passek-Kumericki and J.~Trampetic,
  Phys.{} Rev.{} \textbf{D72}, 054004 (2005)
  [arXiv:hep-ph/0503133];
  B.~Melic, K.~Passek-Kumericki and J.~Trampetic,
  Phys.{} Rev.{} \textbf{D72}, 057502 (2005)
  [arXiv:hep-ph/0507231];
  M.~Buric, D.~Latas, V.~Radovanovic and J.~Trampetic,
  Phys.{} Rev.{} \textbf{D75}, 097701 (2007).

\bibitem{Ohl/Reuter:2004:NCPC}
  T.~Ohl and J.~Reuter,
  Phys.{} Rev.{} \textbf{D70}, 076007 (2004)
  [arXiv:hep-ph/0406098];
  T.~Ohl and J.~Reuter,
  [arXiv:hep-ph/0407337].

\bibitem{Alboteanu/Ohl/Rueckl:2006}
  A.~Alboteanu, T.~Ohl and R.~R\"uckl,
  Phys.{} Rev.{} \textbf{D74}, 096004 (2006)
  [arXiv:hep-ph/0608155];
  PoS \textbf{HEP2005} (2006), 322
  [arXiv:hep-ph/0511188].

\bibitem{Barnich:2003wq}
  G.~Barnich, F.~Brandt and M.~Grigoriev,
  Nucl.{} Phys.{} \textbf{B677}, 503 (2004)
  [arXiv:hep-th/0308092].

\bibitem{Zeiner/Rauh:Theses}
  J.~Rauh,
  Diploma Thesis, University of W\"urzburg, 2006;
  J.~Zeiner,
  PhD Thesis, University of W\"urzburg, 2007.

\bibitem{Ohl/Rauh/Rueckl/Zeiner:2007:AOSWM}
  T.~Ohl, J.~Rauh, R.~R\"uckl, J.~Zeiner, 2007 (in preparation).

\bibitem{Moller:2004qq}
  L.~M\"oller,
  JHEP \textbf{0410}, 063 (2004)
  [arXiv:hep-th/0409085].

\bibitem{Alboteanu:2007}
  A.~Alboteanu,
  PhD Thesis, University of W\"urzburg, 2007.

\bibitem{Asakawa:1999cu}
  T.~Asakawa and I.~Kishimoto,
  JHEP \textbf{9911}, 024 (1999)
  [arXiv:hep-th/9909139].

\bibitem{Haag/Ruelle}
  R.~Haag,
  Phys.{} Rev.{}  \textbf{112}, 669 (1958);
  H. J. Borchers, Nuovo Cim.{} \textbf{25}, 270 (1960);
  D. Ruelle, Helv.{} Phys.{} Acta \textbf{35}, 34 (1962).

\bibitem{CCWZ}
  S.~R.~Coleman, J.~Wess and B.~Zumino,
  Phys.{} Rev.{}  \textbf{177}, 2239 (1969);
  C.~G.~Callan, S.~R.~Coleman, J.~Wess and B.~Zumino,
  Phys.{} Rev.{}  \textbf{177}, 2247 (1969).

\bibitem{Weinberg:1978kz}
  S.~Weinberg,
  Physica A \textbf{96}, 327 (1979).

\bibitem{Wulkenhaar:SWM-theta-expansion}
  J.~M.~Grimstrup and R.~Wulkenhaar,
  Eur.{} Phys.{} J.{}  C \textbf{26}, 139 (2002)
  [arXiv:hep-th/0205153].

\bibitem{Vermaseren:FORM}
  J.{} A.{} M.{} Vermaseren,
  \textit{Symbolic Manipulation with \texttt{FORM} version 2:
    Tutorial and Reference Manual},
  CAN (Amsterdam, 1992).

\bibitem{Alboteanu/Ohl/Rueckl:2007:Pheno}
  A.~Alboteanu, T.~Ohl and R.~R\"uckl, 2007 (in preparation).

\end{thebibliography}
\end{document}